\documentclass[conference]{IEEEtran}
\usepackage{flushend}
\usepackage{amssymb}
\usepackage{graphicx}
\usepackage{epstopdf}
\usepackage{multirow}
\usepackage{amsmath}
\usepackage{subfig}
\usepackage{array}
\usepackage{url}
\usepackage{mwe}
\usepackage[export]{adjustbox}

\IEEEoverridecommandlockouts
\begin{document}
%
\title{Accelerating Viterbi Algorithm using Custom Instruction Approach\thanks{The final publication is available at http://ieeexplore.ieee.org.}}

\author{
\IEEEauthorblockN{Waqar Ahmad\IEEEauthorrefmark{1} and Imran Hafeez Abbassi\IEEEauthorrefmark{2}}
\IEEEauthorblockA{\IEEEauthorrefmark{1}Electrical and Computer Engineering\\
	Concordia University, Montr\`eal, Canada\\
	Email: waqar@encs.concordia.ca\\
\IEEEauthorrefmark{1}\IEEEauthorrefmark{2}School of Electrical Engineering and Computer Science,\\ National University of Sciences and Technology,\\
Islamabad, Pakistan\\
Email: \{waqar.ahmad,imran.abbasi\}@seecs.nust.edu.pk}
\and
\IEEEauthorblockN{Usman Sanwal\IEEEauthorrefmark{3} and Hasan Mahmood\IEEEauthorrefmark{4}}
\IEEEauthorblockA{\IEEEauthorrefmark{3}Computational Biomodeling Laboratory,\\ Department of Computer Science,
\AA bo Akademi University,\\ Turku Centre for Computer Science, 20500 Turku, Finland\\
Email: msanwal@abo.fi\\
\IEEEauthorrefmark{4}Department of Electronics,
Quaid-i-Azam University\\
45320, Islamabad, Pakistan\\
Email: hasan@qau.edu.pk}}


%


\maketitle

\begin{abstract}
In recent years, the decoding algorithms in communication networks are becoming increasingly complex aiming to achieve high reliability in correctly decoding received messages. These decoding algorithms involve computationally complex operations requiring high performance computing hardware, which are generally expensive. A cost-effective solution is to enhance the Instruction Set Architecture (ISA) of the processors by creating new custom instructions for the computational parts of the decoding algorithms.
 In this paper, we propose to utilize the custom instruction approach to efficiently implement the widely used Viterbi decoding algorithm by adding the assembly language instructions to the ISA of DLX, PicoJava II and NIOS II processors, which represent RISC, stack and FPGA-based soft-core processor architectures, respectively.
By using the custom instruction approach, the execution time of the Viterbi algorithm is significantly improved by approximately 3 times for DLX and PicoJava II, and by 2 times for NIOS II.
\end{abstract}


%
\IEEEpeerreviewmaketitle

\section{Introduction}
\label{sec:intro}
In the past few years, there has been a continuous increase in the demand of efficient and reliable transmission of messages over the band-limited and noisy communication channels especially for Internet and wireless networks. In order to meet this demand, many decoding algorithms, such as  Viterbi \cite{forney1973viterbi}, have been developed and enhanced over the years. The Viterbi algorithm~\cite{forney1973viterbi} is one of the most widely used decoding algorithm, which utilizes the Maximum-Likelihood Decoding (MLD)~\cite{forney1974convolutional} procedure in order to reliably decode the transmitted messages at the receiver end. According to an estimate, $10^{15}$ bits/sec are decoded every day by Viterbi algorithm in digital TV devices~\cite{forney2005viterbi}. In Viterbi algorithm, the operations add, compare and select (ACS) have been called many times during the decoding process. This makes the Viterbi algorithm extensively iterative and computationally complex, so it is important to implement this algorithm in a most efficient manner to improve its performance.

Several methods have been proposed in order to implement the Viterbi algorithm efficiently ranging from DSPs~\cite{wilson2003efficient} to FPGA-based \cite{cholan2012design} dedicated hardware designs. However, these implementations have been mainly intended for powerful but expensive high-end DSPs and FPGA devices. A cost-effective solution is to implement the Viterbi algorithm by using the custom instruction approach~\cite{yazdanbakhsh2014implementation}, which is a method of enhancing the ISA of the processors by adding new instructions in order to significantly reduce the execution time of the Viterbi algorithm.

In custom instruction approach, firstly the most computational part of the given algorithm is identified and then new instructions, which implement the identified computational part, generally at the microarchitecture level of the processor, are added to the ISA of the processors. Thus, enabling the modified processors to execute the computationally complex algorithm, such as Viterbi~\cite{forney1973viterbi}, in a most efficient manner compared to its implementation without custom instruction. This approach has been successfully utilized to improve the execution time of many cryptographic algorithms~\cite{ahmed2011efficient,chen2010implementing}, video coding standard~\cite{gonzalez2013acceleration} and  trigonometric functions~\cite{lin2013implementation}.


In this paper, we utilize the above-mentioned custom instruction approach to efficiently implement the Viterbi decoding algorithm in DLX~\cite{hennessy2011computer}, PicoJava II~\cite{tanenbaum2016structured} and NIOS II~ \cite{niosii_proc} representing RISC, stack and FPGA-based soft processors, respectively. DLX and PicoJava II provide microprogrammed control store enabling modification as well as the inclusion of new custom instructions to their ISAs~\cite{hennessy2011computer,tanenbaum2016structured}. We utilize the microprogramming technique to design the custom instructions (\textit{Texpand}) for Viterbi ACS, which are add, compare and select operations that are extensively called during the decoding process to correctly find out the transmitted codeword, and include it in the ISA of DLX and PicoJava II processors.  We then test the performance of the \textit{Texpand} instructions by implementing the ISA of DLX and PicoJava II in CPU\textit{Sim}~\cite{skrien2001cpu} and  MIC-1~\cite{tanenbaum2016structured} simulators, respectively. Currently, the custom instruction based implementation of Viterbi algorithm on DLX is only presented for 12 bits decoding~\cite{ahmed2013efficient}. However, in this paper, we provide the results for up to 60 bits, which can be easily extendable to more number of bits. Since NIOS II is a FPGA-based soft processor, the custom instruction is a dedicated hardware circuitory, which is attached to the NIOS II ALU and invoked when custom instruction is executed. We create the \textit{Texpand} custom instruction in NIOS II by using Verilog HDL programming language~\cite{palnitkar2003verilog} and test it on ALTERA DE2 board CYCLON II FPGA \cite{Quartus} using NIOS II IDE software. We compare the performance of the Viterbi algorithm implementation with and without custom instruction in terms of clock cycle for DLX, PicoJava II and NIOS II processors. The proposed custom instruction approach shows significant improvement in the Viterbi algorithm execution time of $\approx$3 times for DLX and PicoJava II processors, and $\approx$2 times for NIOS II processor. To the best of our knowledge, this is the first work that utilizes the custom instruction approach and presents an efficient implementation of  Viterbi algorithm efficiently in DLX, PicoJava II and NIOS II representing three different processor architectures, i.e., RISC, stack and FPGA-based soft-core processor architectures, respectively.


\section{Related Work}
\label{sec:rel_work}

Viterbi decoding algorithm has been extensively implemented in DSPs and FPGAs in order to reduce the computational complexity. For instance,  Cholan described the FPGA-based design of the Viterbi decoding algorithm and presents an implementation of the decoder for the UWB MB\_OFDM technology \cite{cholan2012design}. Similarly, Ou et. al presented an FPGA-based Viterbi decoder architecture that can provide various throughput and energy trade-offs with an improvement of up to 26.1\% \cite{ou2005time}. Wilson described  an  efficient  implementation  of  the  Viterbi
decoding algorithm on the ZSP500 digital signal processor (DSP)
core~\cite{wilson2003efficient}. However, the state-of-art DSPs and FPGAs are generally expensive and their utilization  for Viterbi implementation may not be a suitable choice for cost-limited applications, such as digital TVs~\cite{saito1992digital}.

The custom instruction approach has been extensively utilized as a cost-effective solution for the efficient implementation of computationally complex algorithms. For instance,  Chen et. al utilized the custom instruction approach to protect cryptographic software implementations against Side Channel Attack (SCA) by emulating the behavior of the secure hardware circuits~\cite{chen2010implementing}. Similarly, custom instruction approach has also been used to efficiently implement the face detection algorithm~\cite{edwards2017real} and S8 AES algorithm~\cite{ahmed2011efficient}. Viterbi algorithm has also been implemented in Xtensa~\cite{Xtensa_convol_coding} and in DLX~\cite{ahmed2013efficient} processors using custom instruction approach. However, in~\cite{Xtensa_convol_coding}, the implementation is described using C programming, which may not be optimized causing extra assembly instructions overhead. Whereas, \cite{ahmed2013efficient} presented the custom instruction based implementation of Viterbi algorithm in DLX processor for 12 bits only.

In this paper, we utilize the custom instruction approach and describes the efficient implementation of Viterbi decoding algorithm up to 60 bits in DLX, PicoJava II and NIOS II processors representing RISC, Java and FPGA-based soft processor architectures, respectively.

\section{Processor Architectures}
\label{sec:proc_arch}
Many processors have hardwired control units composed of digital logic components. The ISA of these processors consist of fixed number of instructions that cannot be modified. On the other hand, there are some processors, such as DLX and PicoJava II, that have microprogrammed control units offering the ability to enhance and modify their ISAs. Similarly, FPGA-based soft core processors, such as NIOS II, also provide the flexibility of adding new custom instructions to their ISAs. In this work, we utilize DLX, PicoJava II and NIOS II processors for accelerating the Viterbi algorithm using custom instruction approach. A brief description of their architectures are described in this section.

\subsection{DLX Processor}
The DLX processor~\cite{hennessy2011computer} has 32 general-purpose registers (R0-R31) of 32 bits. Some registers have special roles. For instance, the value of register R0 is always zero while the branch instructions to subroutines implicitly use register R31 to store the return address. DLX processor memory is byte-addressable and divided into words of 32 bits. Microprogramming consisting of microinstructions have been typically used to derive the DLX datapath.  Some of the commonly used DLX assembly instructions with their microinstructions are shown in Table~\ref{table:dlx_inst}.


 CPU\textit{Sim}~\cite{skrien2001cpu} is a Java-based simulator allowing users to design processors at the microcode level and to run machine-language or assembly-language programs on those processors through simulation. It provides several interesting features to design variety of architectures, including accumulator-based, RISC-like, or stack-based (such as the JVM) architectures. In this paper, we utilize the  CPU\textit{Sim}  simulator to design a DLX processor ISA and then include the \textit{Texpand} custom instruction in order to accelerate the Viterbi algorithm by implementing the microprogramming code for each individual instruction.
\begin{table}[h!]
\caption{DLX Instructions with their Microinstructions}\label{table:dlx_inst}
\center
\begin{tabular}{l l}
\hline\noalign{\smallskip}
DLX  Instruction & Microinstruction \\
\noalign{\smallskip}
\hline
\noalign{\smallskip}
LD R4,100(R1)   & ir(8-15) -$>$ mar\\
		     & Main [mar] -$>$ mdr\\
		     & mdr -$>$ ir(5-7)\\
		     & end\\
SW R4,100(R1) & ir(8-15) -$>$ mar\\
                          & ir(5-7) -$>$ mdr\\
		     & mdr -$>$ Main[mar]\\
		     & end\\
AND R1,R2,R3  & Ir(8-10) -$>$  B\\
		    & Ir(11-13) -$>$ A\\
		    & acc $<$- A \& B\\
		    & acc -$>$ ir(5-7)\\
		    & end\\

\hline
\end{tabular}
\end{table}

%

\subsection{PicoJava II Processor}

PicoJava II~\cite{tanenbaum2016structured} is a 32-bit pipelined stack-based processor, which can execute the Java Virtual Machine (JVM) instructions. There are about 30 JVM instructions that are microprogrammed and typically execute in a single clock.

The instructions in PicoJava II is executed in six pipeline stages. The \textit{first} stage is the instruction \textit{fetch} stage, which takes instructions from instruction cache (I-cache). The \textit{second} and \textit{third} stages are the \textit{decode and fold} stages. The opcode and three register fields are decoded in the decode stage. In the fold stage, the instruction folding operation is performed, in which a particular sequence of instructions is detected and combined into one instruction~\cite{tanenbaum2016structured}. In the \textit{fourth} stage, the operands are fetched from the stack, i.e., from the register file, which are then ready for the \textit{fifth} stage known as the \textit{execution} stage. The results are stored in the cache during \textit{sixth} stage. Some of the PicoJava II assembly language instructions and their microcode are shown in the Table~\ref{table:jvm_inst}. Further detail about the PicoJava II microinstructions and their execution stages can be found in~\cite{tanenbaum2016structured}.

\begin{table}[h!]
\caption{PicoJava II Instructions and their Microinstructions}\label{table:jvm_inst}
\center
 {
\begin{tabular}{p{0.8cm} p{3.5cm} p{3.5cm}}
\hline \hline \noalign{\smallskip}
Mnemonic&\ \ \ Microcode&Description  \\
\noalign{\smallskip}
\hline \hline
\noalign{\smallskip}
iadd1   &MAR = SP = SP - 1; rd&Read in next-to-top word on stack.\\\hline
iadd2   &H = TOS&H = top of stack\\ \hline
iadd3	&MDR = TOS = MDR+H; wr; goto (MBR1)&Add top two words; write to new top of stack\\ \hline
iload1&H = LV&MBR contains index; copy LV to H\\ \hline
iload2&MAR = MBRU + H; rd&MAR = address of local variable to push\\ \hline
iload3&MAR = SP = SP + 1&SP points to new top of stack; prepare write\\ \hline
\end{tabular}
}
\end{table}

\subsection{NIOS II Soft Processor}
\label{sec:nios_proc}
 The basic architectural diagram of the Cyclon II FPGA, designed by ALTERA ~\cite{Quartus}, consisting of several peripherals, such as SDRAM, SRAM and UART, and their interface with NIOS II processor through Avalon Switch Fabric~\cite{niosii_proc}. NIOS II processor need interfaces to connect to other devices on the board, that are instantiated on the Cyclon II FPGA chip along with the NIOS II processor. These interfaces are connected to each other by means of a interconnection network known Avalon Switch Fabric. In this network, the master components are on one side and slave component are on the other side. The key responsibility of the Avalon Switch Fabric is to synchronize the transfer of data between two devices.



NIOS II soft processor  is available in three different versions, i.e., economy (e), standard (s) and fast (f) processors \cite{niosii_proc}. All of these processors have separate instruction and data caches except NIOS II/e. About $256$ custom instructions can be added to the ISA of these NIOS II processors. NIOS II processors can be created in ALTERA DE2 board~\cite{altera_de2_board} using the SOPC builder in Quartus II software \cite{Quartus}. By using the \textit{Add new component} feature in SOPC builder, we can add new custom instructions in the ISA of NIOS II processor \cite{niosii_proc}. The custom logic is then attached to the NIOS II ALU and is invoked when custom instruction is executed.


\section{Viterbi Algorithm}
\label{sec:viterbi}
\hskip 0.22in
A typical communication system incorporates channel coding schemes in order to correct transmission errors. The process of channel coding involves the addition of redundancy in the information bits. Over the years, many channel coding schemes have been developed, which are mainly distinguished by their error correcting capabilities against channel noise. There two major types of codes, i.e., Block or Convolutional, which are differ by their encoding  principle.  In Block Codes, the information bits are followed by the parity bits while the later convolve the sequence of information bits to codewords sequentially according to some specified rules. Viterbi algorithm has been extensively utilized for decoding both types of codes~\cite{bossert1999channel}. However, in this paper, we mainly focus on the convolutional codes that are generated from the convolutional encoder. The encoding process of information bits to codeword using the convolutional encoder is briefly described in the next section.



\subsection{Encoding}
\label{subsec:encod}

\begin{figure}[!htb]
	\centering
	\centering
	\subfloat[]{\includegraphics[scale=0.45]{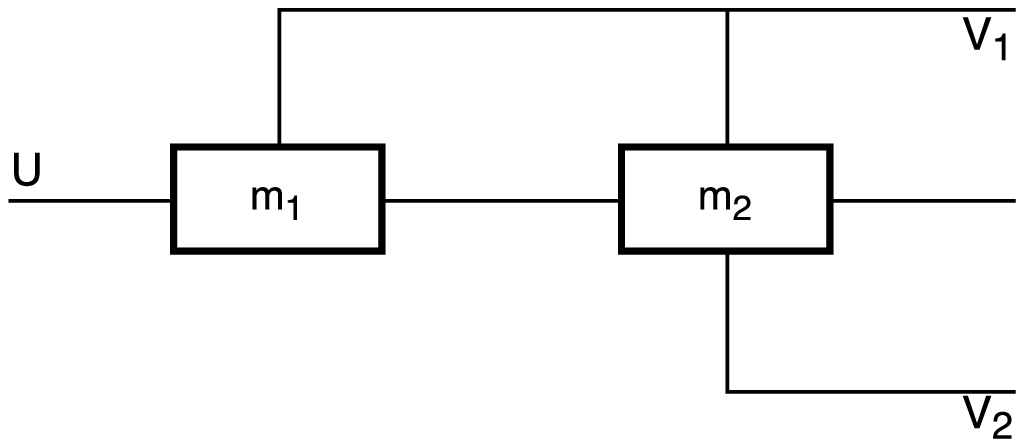}} \qquad
	\subfloat[]{\includegraphics[scale=0.2]{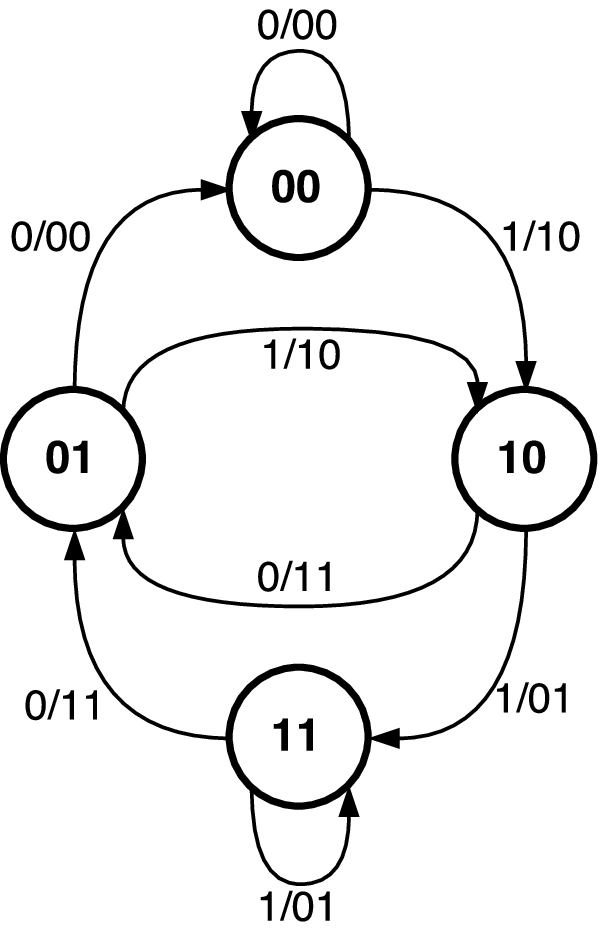}}\vspace{8pt}
	\newline
	\subfloat[]{\includegraphics[scale=0.45]{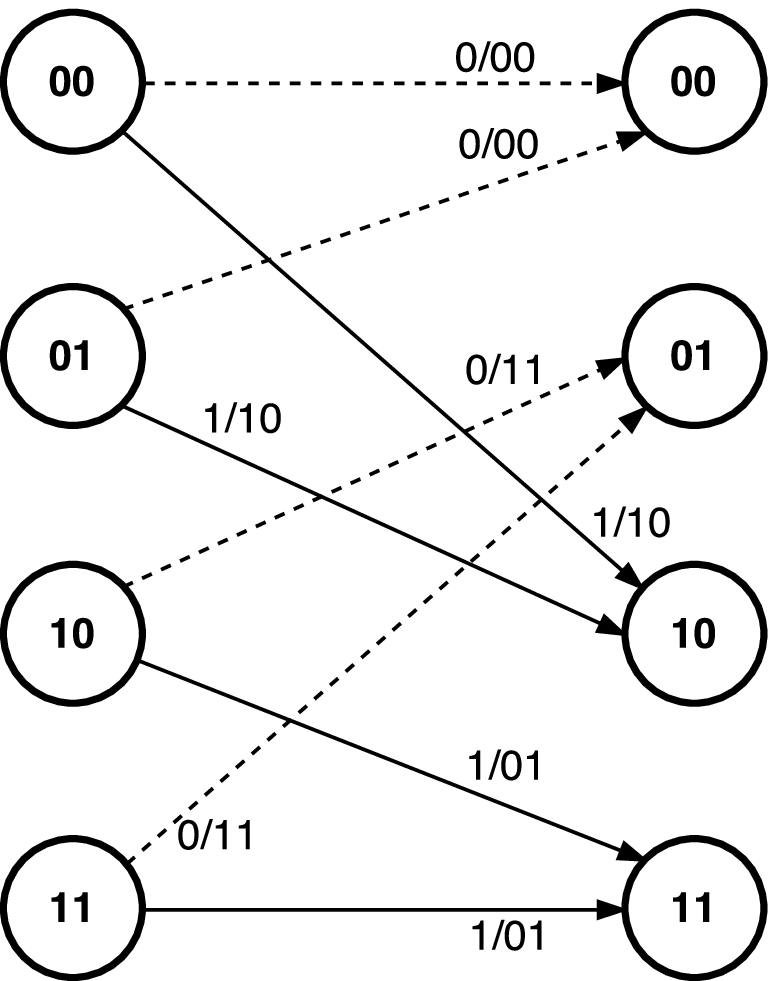}}
	\caption{(a) State diagram of convolutional encoder, (b) A typical convolutional encoder (c) Trellis for convolutional encoder} \label{convol_encod}
\end{figure}

 Convolutional encoders are discrete-time linear time-invariant (LTI) systems that have been typically used to encode $K$ information  bits  to generate $N > K$ codewords  in  each  time  step~\cite{bossert1999channel}.  A convolutional encoder having coding rate $\frac{1}{2}$ is shown in Figure~\ref{convol_encod}(a), where $U$ represents the information bits and $V_1$ and $V_2$ are the corresponding output generated by the encoder from each information bit in a sequential manner. The memory elements $m_1$ and $m_2$ represent state of the encoder during the encoding of information bits.


 A complete state diagram of the convolution encoder is shown in Figure~\ref{convol_encod}(b). The nodes represent the state of the encoder whereas the edges describe the transitions between states based on the input/output relationship of the information bits with the code bits. A state diagram can be equivalently represented in the form of a trellis diagram, as shown in Figure~\ref{convol_encod}(c). The trellis is a special graph with edges representing the possible transitions from states and is considered as the backbone in the decoding process of the convolutional codes.

 In order to illustrate the working of the convolutional encoder, as shown in Figure~\ref{convol_encod}(b), consider an information bits (110100) having first four bits are data bits while the last two are flush bits. After passing the information bits through the encoder, the resulting codeword bits are (10 01 11 10 11 00). Assuming, if the noisy channel caused the $3^{rd}$ and $7^{th}$ bits of codeword in error then the received codeword becomes (10 11 11 00 11 00).

\subsection{Decoding}

\begin{figure}[h!]
	\center
	\includegraphics[trim={0 2cm 0 0},clip,scale=0.3]{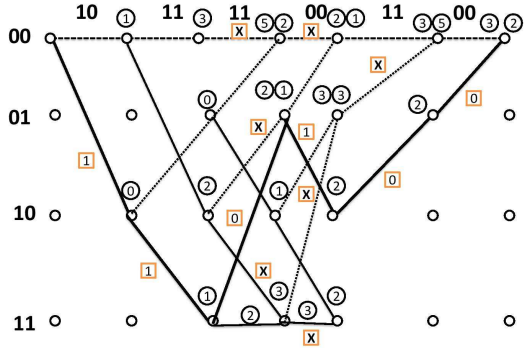}
	\caption{Trellis Diagram for Typical Application}\label{trellis_diag}
\end{figure}
Viterbi algorithm utilizes the trellis structure to perform the decoding operation. The number of times the trellis expansion function is called depends upon the amount of decoding bits and the states in the trellis. Based on the above design, the complete trellis diagram is shown Fig.~\ref{trellis_diag}. This trellis diagram describes a way to select the path with minimum weight among all the paths. In Fig.~\ref{trellis_diag}, the recieved bits are shown at the top of trellis diagram whereas in the left side corner all the possible states of the encoder are listed. The trellis expands from state (00) and only those paths survive which end at the state (00). The dashed lines are the paths that originates when $0$ bit is given as input to the encoder whereas the solid lines are obtained from $1$ input bit to the encoder. The path which has the minimum weight among all the surviving paths is shown by a dark solid line.

In Viterbi algorithm, after a transition from a state to the next state, the weights are calculated for each possible path. The path weight is incremented whenever there is a difference in a particular received bit and the state output bit in the transition path. For example, if the received bits are (00) and output bit in a particular state transition is (01), then the path weight value is incremented by 1 for a particular  path transition. Similarly, in the case of difference of two bits the path weight is incremented by 2. If more than one paths arrive at a particular state, the path with lowest weight survives and the remaining paths are deleted. For the case when the weights of the arriving paths are equal, the path arriving from the lowest state survives. For instance, if we have state (00) and state (01) both arrive at state (00) having same path weight values, we select the path that arrives from state (00). When the decoding process is completed, a trace back function is performed to determine the most probable transmitted sequence by selecting the path that start from state (00) and ends up at state (00) having minimal path weight among all the paths. In Fig. \ref{trellis_diag}, the weight of the path at each node is represented in a circle whereas the square-box shows the correctly decoded information bits. The cross in square-box represents that the corresponding path does not survive.

\subsection{Custom Trellis Instruction}
\label{subsec:custom_trellis}

From the last section, it is evident that the trellis expansion process, involving add, compare and select (ACS) operations, are called several times as we progress in the decoding process. For example, if there are 12 bits in a received codeword then this trellis expansion function is called almost 19 times. Therefore, it is desirable to create a custom trellis expand instruction that allows the processors to execute these operations in a minimum clock cycles in order to achieve maximum efficiency.

We create \textit{Texpand} custom instructions in DLX, PicoJava II and NIOS II processors performing two fundamental tasks. The \textit{first} task is the implementation of the following operations: (i) add- the calculation of the cumulative weights of the arriving paths at a particular state; (ii) compare- a comparison operation between the weights of the arriving paths; and (iii) select- a selection operation to find out the surviving path. The \textit{second} task is to keep track of the path with minimum weight that ultimately ends up at state (00). At the final stage, the trace-back function is performed, based on the path with minimum weight, in order to determine the most probable transmitted sequence.

\section{Comparison}
\label{sec:comp}
In this section, we present a performance comparison between trellis expansion function, which is written in assembly language, and the \textit{Texpand} custom instruction that is created and included in the ISA of DLX by using CPU\textit{Sim}, PicoJava II by using Mic-1 and in NIOS II by using SOPC builder. Each microinstruction in DLX and PicoJava typically takes 4 clock cycles to complete its execution. The comparison is made on the basis of number of clock cycles consumed by the microinstructions in the implementation of trellis expansion function and the Texpand custom instruction. Viterbi algorithm is implemented for 12 bits decoding having trellis expansion function as well as Texpand instruction is called about 19 times. Tables~\ref{table:DLX} and~\ref{table:picojava} show a significant performance improvement of about 3.5 and 3 times for DLX and PicoJava II processors, respectively.

\begin{table}[h!]
\caption{Comparison between trellis assembly function and Texpand Instruction on CPUSIM}\label{table:DLX}
\centering
\resizebox{\linewidth}{!} {
\begin{tabular}{l l | l l}
\hline
\hline
\noalign{\smallskip}
Trellis Assembly Function&  & Texpand Instruction\\
\noalign{\smallskip}
\hline
\hline
\noalign{\smallskip}
Assembly Instruction (A.I)	 & 63 &      Assembly Instruction (A.I) & 1\\ \hline
Microinstruction (M.I)	 & 277 & Microinstruction (M.I) & 100\\ \hline
Fetched Instructions (I x 4)			& 63 & Fetched Instructions (I x 4)	 & 1\\ \hline
Function calls	        & 19 & Texpand instruction calls & 19\\ \hline
Total M.I =   		       & 6460 & Total M.I =  		       & 1919 \\
((M.I + F.I) x 19)	       &  & ((M.I + F.I) x 19)	       &  \\ \hline
Total Time (T) = M.I x 4			&  25840 & Total Time (T) = M.I x 4			& 7676\\
			 & & \%age Improvement & 236\\
\hline
\end{tabular}
}
\end{table}

\begin{table}[htb!]
\caption{Comparison between Trellis assembly function and Texpand Instruction on MIC-1} \label{table:picojava}
\center
\resizebox{\linewidth}{!} {
\begin{tabular}{l  l | l l}
\hline
\hline
\noalign{\smallskip}
Trellis Assembly Function&  & Texpand Instruction\\
\noalign{\smallskip}
\hline
\hline
\noalign{\smallskip}
Assembly Instruction (A.I)	 & 41 &      Assembly Instruction (A.I) & 1\\ \hline
Microinstruction (M.I)	 & 255 & Microinstruction (M.I) & 102\\ \hline
Fetched Instructions (I x 4)		& 41 & Fetched Instructions (I x 4) & 1\\ \hline
Function calls	        & 19 & Texpand instruction calls & 19\\ \hline
Total M.I   		       & 5624 & Total M.I   		       & 1957 \\
((M.I + F.I) x 19)	       &  & ((M.I + F.I) x 19)	       &  \\ \hline
Total Time (T) = M.I x 4				&  22496 & Total Time (T) = M.I x 4			& 7828\\
			 & & \%age Improvement & 187 \\
\hline
\end{tabular}
}
\end{table}

Similarly, TABLE~\ref{table:nios} shows the comparison between the NIOS II assembly language program with custom instruction-based Viterbi algorithm implementation. The execution of the NIOS II assembly instructions take different clock cycles and its also different for each version~\cite{niosii_proc_core_implem}.  The performance of the Viterbi algorithm is considerably improved by about 2 times with the custom instruction as compared to the assembly language function for all the NIOS II processors. Some additional assembly instructions are used in the assembly program with custom instruction because the data that are required to pass to the custom instruction is embedded in the register through shift instruction before calling the custom instruction. After the execution of the custom instruction, the results in the particular register can only be extracted by using additional assembly instructions.

\begin{table*}[!htb]
	\caption{Comparison between NIOS II Trellis assembly function and Custom Instruction} \label{table:nios}
	\centerline {\small A.L.T.F = Assembly Language Trellis Expansion Function,  C.I = Custom Instruction,} \centerline{\small A.L.I = Assembly Language Instructions}
	
	\center
	\begin{tabular}{l  l | l  l | l  l }
		\hline \hline\noalign{\smallskip}
		Nios II/f Processor & & Nios II/s Processor & & Nios II/e processor\\
		\noalign{\smallskip}
		\hline \hline
		\noalign{\smallskip}
		\# of cycles used in A.L.T.F	& 59 & \# of cycles used in A.L.T.F	& 59 & \# of cycles used in A.L.T.F	& 264\\
		Function calls	& 19 & Function calls	& 19 & Function calls	& 19\\
		 Total no. of cycles used in A.L.T.F	& 1121 &  Total no. of cycles used in A.L.T.F	& 1121 & Total no. of cycles used in A.L.T.F	& 5016\\ \hline
		\# of cycles used in C.I + A.L.I	& 28 & \# of cycles used in C.I + A.L.I	& 35 & \# of cycles used in C.I + A.L.I	& 151\\
		Texpand Instruction calls	& 19 & Texpand Instruction calls	& 19 & Texpand Instruction calls	& 19\\
		Total no. of cycles used with C.I + A.L.I  & 532 &  \# cycles used in C.I +   A.L.I   & 665 & \# cycles used in C.I +  A.L.I  & 2869\\
		Performance Improvement		& 110.7\% & Performance Improvement		& 68.5\%  & Performance Improvement		& 74.8\%\\
		\hline
	\end{tabular}
\end{table*}

As it can be seen from Tables~\ref{table:DLX}, \ref{table:picojava}, and \ref{table:nios} that the performance improvements of Viterbi algorithm in DLX and PicoJava II processors are quite higher then its performance improvements on NIOS II soft processors. For DLX and PicoJava II implementations, there is no need to use the shift instructions to pass the data to the custom instruction, we can directly access the data in the memory location through microinstructions in CPU\textit{Sim} and MIC-1 simulators while in NIOS II the data are passed to the custom instruction by using additional assembly instructions. Therefore, additional execution cycles are consumed in the assembly program of Viterbi algorithm with custom instruction. Another reason is that the technique of custom instruction in NIOS II is based on writing a Verilog HDL program, which is quite different than the procedure used in the creating custom instructions CPU\textit{Sim} and MIC-1 simulators.

%
%

\begin{figure}[!htb]
	\centering
	\centering
	\subfloat[]{\includegraphics[trim={0 0 0 0},clip, scale=0.35]{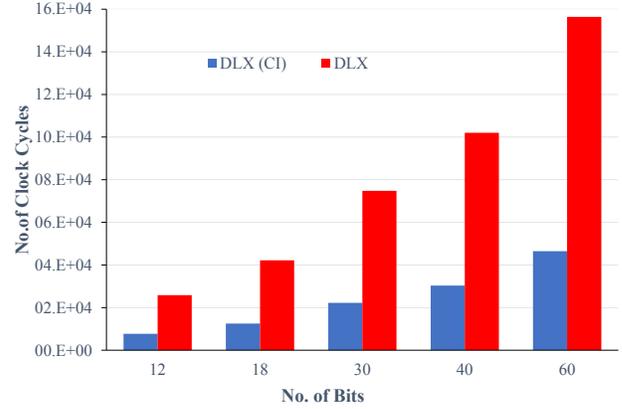}}\\
	\subfloat[]{\includegraphics[trim={0 0 0 0},clip, scale=0.35]{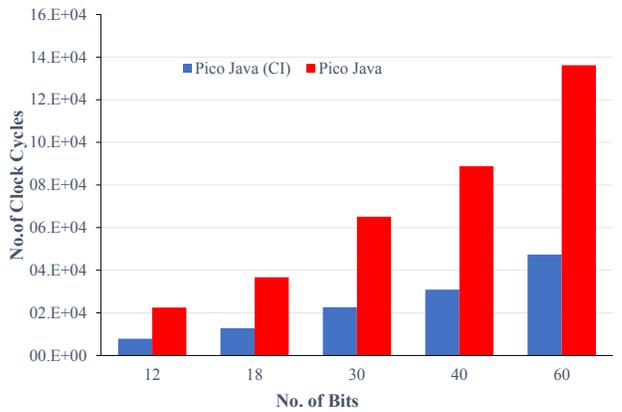}}\vspace{8pt}
	\newline
	\subfloat[]{\includegraphics[trim={0 0 0 0},clip, scale=0.35]{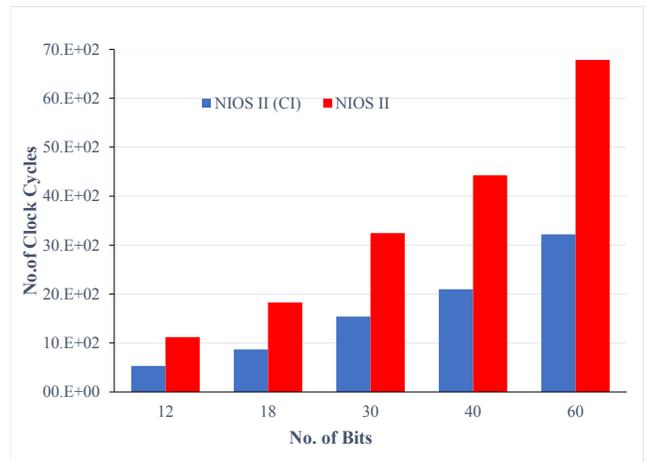}}
	\caption{Histogram Plot for DLX, PicoJava II and NIOS II Performance Improvement}
	\label{fig:dlxhisto}
\end{figure}

\begin{figure}[!htb]
 	\centering
 	\includegraphics[trim={0 0 0 0},clip, scale=0.37]{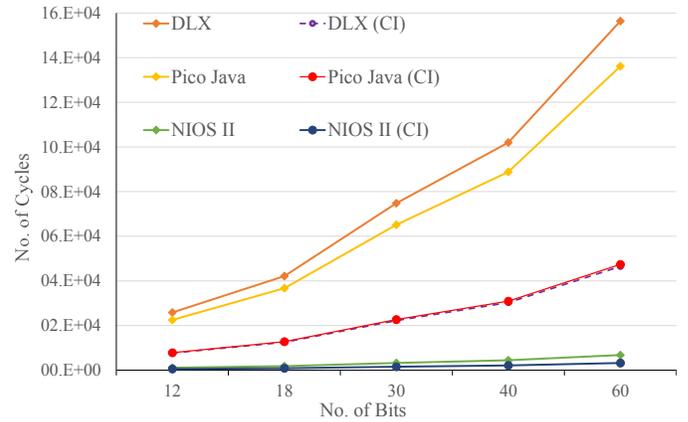}
 	\caption{Trend of Performance Improvements in DLX, PicoJava II and NIOS II}
 	\label{fig:combinegraph}
 \end{figure}

 The trend in the performance improvement of Viterbi algorithm, using our custom Texpand instruction, by increasing the number of received bits,  can be seen in Fig.~\ref{fig:dlxhisto}. The x-axis shows the number of decoding bits and the y-axis represents the number of clock cycles consumed in order to recover the information bits using the Viterbi algorithm. The first bar depicts the total number of clock cycles used with the custom instruction approach whereas the second bar represents the number of clocks utilized in assembly level program without using the custom instructions. We take the clock cycles consumption as the metric of comparison to measure the performances of Viterbi algorithm implementations, i.e., with and without custom instruction. It can be seen clearly, in Fig.~\ref{fig:dlxhisto}, that the clock cycles are less consumed when viterbi algorithm is implemented with custom instruction compared to its implementation using non-modified ISAs of DLX, PicoJava II and NIOS processors. Also, it can be observed that as the number of bits increases the clock cycles consumption in Viterbi algorithm increases drastically and the custom instruction based implementations significantly help to reduce the number of clock cycles.  For a bird-eye view, Fig.~\ref{fig:combinegraph} presents a graphical depiction of the performance improvement of Viterbi algorithm by using the Texpand instructions compared to assembly program for DLX, PicoJava II and NIOS II processors.

 The proposed implementation of Viterbi algorithm is quire efficient then the custom instruction-based Viterbi algorithm implementation  in Xtensa~\cite{gonzalez2000xtensa}, which is also a FPGA-based soft processor, like NIOS II. In \cite{Xtensa_convol_coding}, the comparison and performance improvement is described between the implementation of C language program and the custom instruction, named as TIE. However, the generated assembly code from C program may not be optimized compared to hand-written assembly program causing extra assembly instructions overhead.  Consequently, consuming more number of cycles effecting the overall performance of the processor.

The custom instructions that we have created in this work are generic and can be used for practical convolutional encoders having coding rate 1/2. For instance, the GSM convolutional encoder \cite{grech1999channel}, which has coding rate 1/2 and constraint length $K$ is $5$ and total number of states is $16$. An important feature of our proposed approach is that it can be used to improve the execution performance of other computationally complex algorithms especially that are used in domain of image processing. For instance, Sundararajana et. al have recently implemented a custom instruction based FFT algorithm on NIOS II processor using ALTERA DE2 board embedded with Cyclone II FPGA~\cite{sundararajana2016custom}. By using our proposed approach, a fair comparison of performance improvement of their custom instruction based FFT algorithm implementation can be analyzed by implementing it also on DLX and PicoJava II processors.

\section{Conclusion}
\label{sec:concl}
 In this paper, we report an enhancement in DLX and PicoJava II processor ISA for efficient implementation of Viterbi decoding algorithm.  We create a custom trellis expansion instruction (Texpand) in CPUSIM simulator on RISC based architecture and MIC-1 simulator on stack based architecture. The execution time is stupendously improved to approximately three times, when Texpand instruction is designed for RISC architecture and approximately three times for stack based architecture. In addition, we enhance the ISA of NIOS II soft processor for the efficient implementation of Viterbi algorithm. The comparison with and without the custom instruction shows substantial improvement in the results. The performance of the NIOS II processor with the custom instruction is improved to two times to the assembly language program without the custom instruction.

In this paper, we presented our proposed approach by realizing the implementation of DLX and PicoJava II
processors on computer-based software tools. However, an FPGA based implementation of these processors may also improve the execution performance for computationally complex algorithms as we can change the clock frequency and also execute the custom instruction in parallel to other independent instructions. We also plan to extend our proposed approach to state-of-the-art architectures, such as GPU~\cite{che2008accelerating}, and aiming to provide a detailed comparison in terms of execution time, delay, latency and complexity. 

\bibliographystyle{plain}
\bibliography{biblio}

\end{document}